\UseRawInputEncoding

\documentclass[prX,twocolumn,nofootinbib,superscriptaddress]{revtex4}

\usepackage{graphicx}
\usepackage{xcolor}

\usepackage{amsmath,srcltx}
\usepackage{latexsym}
\usepackage{amssymb}
\usepackage{amsfonts}
\usepackage{multirow}

\newcommand{\be}{\begin{equation}}
\newcommand{\ee}{\end{equation}}
\newcommand{\ba}{\begin{eqnarray}}
\newcommand{\ea}{\end{eqnarray}}

\newcommand{\aand}{\;\;\;\mbox{and}\;\;\;}

\def\I{\leavevmode\hbox{\small1\kern-3.8pt\norfeynmpmalsize1}}

\begin{document}
\title{Is there any Nambu monopolium out there?}

\author{D.O.R. Azevedo}
\email{daniel.azevedo@ufv.br}
\affiliation{Universidade Federal de Vi\c cosa (UFV),\\
Departamento de F\'\i sica - Campus Universit\'ario,\\
Avenida Peter Henry Rolfs s/n - 36570-900 -
Vi\c cosa - MG - Brazil.}

\author{M.L. Bispo}
\email{milena.bispo@ufv.br}
\affiliation{Universidade Federal de Vi\c cosa (UFV),\\
Departamento de F\'\i sica - Campus Universit\'ario,\\
Avenida Peter Henry Rolfs s/n - 36570-900 -
Vi\c cosa - MG - Brazil.}

\author{O.M. Del Cima}
\email{oswaldo.delcima@ufv.br}
\affiliation{Universidade Federal de Vi\c cosa (UFV),\\
Departamento de F\'\i sica - Campus Universit\'ario,\\
Avenida Peter Henry Rolfs s/n - 36570-900 -
Vi\c cosa - MG - Brazil.}
\affiliation{Ibitipoca Institute of Physics (IbitiPhys),\\
36140-000 - Concei\c c\~ao do Ibitipoca - MG - Brazil.}

\author{D.H.T. Franco}
\email{daniel.franco@ufv.br}
\affiliation{Universidade Federal de Vi\c cosa (UFV),\\
Departamento de F\'\i sica - Campus Universit\'ario,\\
Avenida Peter Henry Rolfs s/n - 36570-900 -
Vi\c cosa - MG - Brazil.}

\author{A.R. Pereira}
\email{apereira@ufv.br}
\affiliation{Universidade Federal de Vi\c cosa (UFV),\\
Departamento de F\'\i sica - Campus Universit\'ario,\\
Avenida Peter Henry Rolfs s/n - 36570-900 -
Vi\c cosa - MG - Brazil.}

\begin{abstract}
Magnetic monopoles have been a subject of study for more than a century since the first ideas by A. Vaschy and P. Curie, circa 1890. In 1974, Y. Nambu proposed a model for magnetic monopoles exploring a parallelism between the broken symmetry Higgs and the superconductivity Ginzburg-Landau theories in order to describe the pions quark-antiquark confinement states. There, Nambu describes an energetic string where its end points behave like two magnetic monopoles with opposite magnetic charges -- quark and antiquark. Consequently, not only the interaction among monopole and antimonopole, mediated by a massive vector boson (Yukawa potential), but also the energetic string (linear potential) contributes to the effective interaction potential. We propose here a monopole-antimonopole non confining attractive interaction of the Nambu-type, and then investigate the formation of bound states, the {\it monopolium}. Some necessary conditions for the existence of bound states to be fulfilled by the proposed Nambu-type potential, Kato weakness, Set\^o and Bargmann conditions, are verified. In the following, ground state energies are estimated for a variety of {\it monopolium} reduced mass, from $10^2$MeV to $10^2$TeV, and Compton interaction lengths, from $10^{-2}$am to $10^{-1}$pm, where discussion about non relativistic and relativistic limits validation is carried out.
\end{abstract}

\maketitle

{\it Dedicated to the memory of Dr. C\'esar Augusto Parga Rodrigues (1939-2021)}

\section{Introduction}
Since the last decade of the 19th century, the asymmetry in Maxwell's equations has drawn the attention of the scientific community \cite{vaschy}. The perfect interplay between electric and magnetic fields is broken by the introduction of field sources as electric charge carriers. This feature inspired the first inquisitions about the possibilities of particles carrying a magnetic charge, called magnetic monopoles, such as from Curie \cite{curie} in 1894. Then, in 1931, Dirac publishes a seminal work \cite{dirac}, in which he proposes an explanation for the electric charge quantization given the existence of a single magnetic monopole and revisits his interpretation of free states of matter with negative energy, coming up with the idea of anti-matter. Although initially received with disbelief, anti-matter was observed by Anderson \cite{anderson} in 1932.

 As for the magnetic monopole, no measure nor observation has yet presented it unmistakably. Nevertheless, it kept reappearing throughout the 20th century, in different theoretical contexts \cite{cf,thooft,polyakov,polchinski}. Some of the examples are: the proposal of Cabbibo-Ferrari \cite{cf}, which involves the addition of an extra gauge field associated with the monopole; the 't Hooft-Polyakov monopole \cite{thooft,polyakov}, proposed independently by the authors, which is a topological soliton that arises with the spontaneous symmetry breaking in a Yang-Mills theory; in string theory \cite{polchinski}, as shown by J. Polchinski, where it is related to the properties of Dirichlet-branes. In all of these theories, Dirac's quantization condition arise, reinforcing the initial idea, but they all appear as finite, smooth solutions, not resulting in any singularity.
 
The use of magnetic monopoles as an explanation for strong nuclear interactions \cite{schwinger} was another one of these ideas. Since the interaction between two monopoles would be much stronger than the analogous electric interaction, due to Dirac's quantization condition, they were candidates for the role of quarks. One of such works was proposed by Nambu \cite{nambu} in 1974, in which the Dirac description of magnetic monopoles is adapted to a London-type theory, exploring the Nielsen-Olesen interpretation of dual strings. The result is that quarks in this model behave as carriers of magnetic charge and are bound together by the dual string, giving us the meson. Here it is explored the possibility of magnetic monopoles as fundamental particles which behave like the description given by Nambu for the quarks. However, it should be noticed that unlike the long range Dirac string, which is unphysical, so unmeasurable, the short range Nambu string is physical, measurable, thus an energetic string playing a critical role in the interaction process among the monopole (quark) and the antimonopole (antiquark), their confinement into meson. Additionally, bearing in mind the fundamental issues about the differences between the proposals by Dirac and Nambu for describing magnetic monopoles, it becomes evident in Dirac the quantum mechanics (microscopic) to classical (macroscopic) electromagnetism connection, giving rise to the electric charge quantization. On the other side, the strictly quantum (microscopic) nature in Nambu is conjectured -- ``here we are using the words electric and magnetic not to refer to actual electromagnetism, but as an analogy in a simplified model of strong interactions'' \cite{nambu} -- in spite of assuming the Dirac quantization condition as a premise.

Interestingly, magnetic monopoles is an object very investigated in condensed matter physics \cite{Volovik}. Recently they were found as emergent quasi-particles in frustrated magnetic materials known as spin ices \cite{Castelnovo,Mol2009}. They arise as a result of the phenomenon called fractionalization. Particularly, in a two-dimensional synthetic material, known as artificial spin ice (ASI), the emergent magnetic monopoles are very similar to the Nambu monopoles \cite{Mol2009,Silva2013,Morgan2011,Marrows2019}, since the string connecting the opposite poles is energetic. Such Nambu monopoles and their strings can be directly observed in ASI systems \cite{Morgan2011,Marrows2019,Goryca} and, therefore, these materials may provide new technologies in which magnetic charges would be the information carriers (magnetricity and magnetronic). Therefore, this subject is an important topic in both, the low energy and high energy physics. In high energy physics arena, the experimental \cite{macro,balestra} and theoretical \cite{epele2009,epele2012,vento2021} quests for fundamental magnetic monopoles became more intense since the beginning of dark energy and dark matter era. Despite of conjectures about that the magnetic monopoles masses could be so far beyond the energy scale in experiments nowadays, if the {\it monopolium} -- monopole-antimonopole bound states -- exists indeed, these masses should be smaller, than those expected according to grand unified models, thanks to the highly strong monopole-antimonopole interaction \cite{epele2008}.    

Here, our proposal is to investigate the possibility of monopole-antimonopole bound states, the {\it  monopolium}, stemming from a nonconfining version of the Nambu interaction potential, even though by adopting some assumptions of the original work by Nambu, namely the Dirac quantization condition and no classical electromagnetism counterpart due to the short range of the Nambu potential and its merely quantum character. The physical considerations in order to adapt Nambu's description are established in Section II, with some necessary modifications on the interaction potential described in Section III. The possibility of bound states is investigated in Section IV and the ground state energy of the monopolium is estimated in Section V, via the variational method. Section VI is left for the discussion of results, conclusions and perspectives.

\section{Initial considerations}
Nambu \cite{nambu} describes the mesons as two quarks carrying opposite magnetic charges -- a quark and antiquark confining state -- interacting through two massive fields, a scalar and a vector one, leading to a interaction of the type\footnote{In the natural system of units where $\hbar=c=1$.}:
\begin{equation}
\label{nambu-potential}
V(r)= -\frac{g^2}{4\pi}\frac{e^{-m_v r}}{r} +  \frac{g^2}{8 \pi} \left[m_v^2\ln{\left(\frac{m_s^2}{m_v^2}+1\right)}\right]r~,
\end{equation}
where $+g$ ($-g$) is the magnetic charge of the quarks (antiquarks), $m_s$ and $m_v$ the masses of the scalar and the vector bosons, respectively, and $r$ is the distance between the quark and antiquark. It can be seen from Eq.(\ref{nambu-potential}) that the binding energy of the {\it string} is proportional to its length, as expected in dynamics of confining states. Bearing in mind that our aim is to propose magnetic monopoles interacting via a non confining version of the Nambu potential, consequently the Nambu-type potential proposed here has similar structure of the original Nambu model (\ref{nambu-potential}), although not of the same nature. We shall assume that the interaction among the magnetic monopole and the magnetic antimonopole is mediated by a single boson, thus the {\it string} energy term, the second term in (\ref{nambu-potential}), depends only on the mass ($\mu$) of this single (scalar or vector) boson. By adopting the first assumption, namely the single boson mediation, the Nambu-like interaction, is expressed by:
\begin{equation}
\label{nambu-potential-single}
V(r) = \dfrac{g^2}{4\pi}\left(\mu^2 r - \dfrac{e^{-\mu r}}{r}\right)~,
\end{equation}
where $g$ is the absolute value of the monopole and antimonopole magnetic charges, and $\mu$ is the mass of the mediating boson.

The non relativistic dynamics of this two-particle (magnetic monopole and antimonopole) system is described in the center of mass by the Schr\"odinger equation:
\begin{equation}
-\dfrac{1}{2m} \nabla^2 \Psi(\mathbf{r},t) + V(r)\Psi(\mathbf{r},t) = i  \dfrac{\partial}{\partial t}\Psi(\mathbf{r},t)~, \label{main}
\end{equation}
where $\mathbf{r}$ is the two-particle relative position ($|\mathbf{r}|=r$), $m$ is the reduced mass and $\Psi(\mathbf{r},t)$ is the two-particle wave function. Owing to the fact that the interaction potential is central and isotropic, then the wave function can be expressed in spherical coordinates by a radial function $R_n(r)$ and the spherical harmonics $Y_l^m(\theta,\phi)$, such that $\Psi(\mathbf{r},t)=R_n(r)Y_l^{m_l}(\theta,\phi)U(t)$, with $n$ being the principal quantum number ($n=1,2,\dots$), $l$ the orbital quantum number ($l=0,\dots,n-1$) and $m_l$ the magnetic quantum number ($|m_l|=0,\dots,l$). Also, as we are searching for possible monopole-antimonopole bound states, ({\it monopolium}), the time dependence is simply $U(t)=e^{-iE_n t}$, with $E_n$ being the energy of a particular bound state associated to $n$. Finally, time-independent radial Schr\"odinger equation reads:
\begin{equation}
\label{radial}
\frac{d^2u(r)}{dr^2} - 2m\left[V(r) + \frac{l(l+1)}{2mr^2}\right]u(r) = E_n u(r)~,
\end{equation}
where $u(r)=rR_n(r)$.

\section{The interaction potential}
\begin{figure}[h]
\centering
\setlength{\unitlength}{1,0mm}
\includegraphics[width=6.05cm,height=5.5cm]{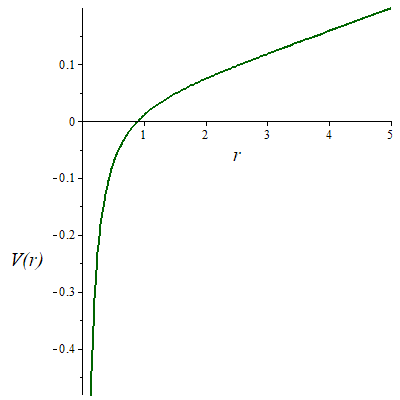}
\caption[]{The monopole-antimonopole potential $V(r)$ (\ref{nambu-potential-single}), with $g=\mu=1$}
\label{nocut}
\end{figure}

At first sight, the potential (\ref{nambu-potential-single}) seems to present no problems in describing monopole and antimonopole interacting. However, it should be noted that the interaction potential (\ref{nambu-potential-single}) is described by two pieces, in the first the binding energy increases linearly with the distance between the two particles while the second represents an Yukawa-type binding energy, {\it i.e.} a short-range interaction. Furthermore due to the part of the potential (\ref{nambu-potential-single}) which behaves linearly with the distance, as the monopole and antimonopole get apart from each other, the interaction energy gets higher (see Fig.\ref{nocut}), describing a confining potential indeed, thus for that reason those monopole and antimonopole would never exist as free particles.

Taking into consideration that one of the main goal is to study monopole-antimonopole bound states and not confining states, some conditions might be imposed onto the interaction potential (\ref{nambu-potential-single}), particularly regarding its linear behavior, so as to allow the existence of monopoles and antimonopoles as free one-particle states. To switch the confining behavior of the {\it string} potential to a non confining, one can restrict the {\it string} to a maximum length, thus bounding the {\it string} energy from above. By establishing that the {\it string} maximum length is $r_0$, in such a way that $V(r_0)=0$:
\begin{equation}
\dfrac{g^2}{4\pi}\left(\mu^2 r_0 - \dfrac{e^{-\mu r_0}}{r_0}\right) = 0~,
\end{equation}
whence the solution $r_0\equiv r_0(\mu)$ of the above transcendental equation reads
\begin{equation}
\label{r0}
r_0 = \frac{2}{\mu}~ W\left(\dfrac{1}{2}\right)~,
\end{equation}
where $W$ is the principal Lambert function.

\begin{figure}[h]
\centering
\setlength{\unitlength}{1,0mm}
\includegraphics[width=6.05cm,height=5.5cm]{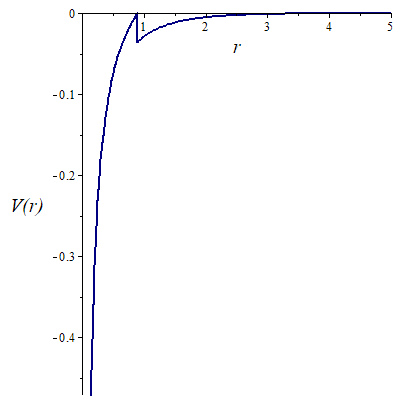}
\caption[]{The potential after {\it deconfinement} displaying discontinuity at $r=r_0$, with $g=\mu=1$}
\label{descontinuo}
\end{figure}

At this moment, it shall be stressed that by carrying out such a {\it deconfinement} of the potential (\ref{nambu-potential-single}), through a length cutoff ($r_0$) in its linear dependence, it becomes discontinuous at $r=r_0$ (see Fig.\ref{descontinuo}). In the case of discontinuity in the potential, it could jeopardize ground and excited states stability, by yielding metastable two-particle states. Accordingly, in order to avoid monopole-antimonopole metastable states, continuity of the potential is required, which can be accomplished by redefining the Nambu-type monopole-antimonopole interaction potential as below:
\begin{equation}
\label{nambu-type-potential}
    V_{\rm N}(r) = \frac{g^2}{4\pi}
    \left\{
	\begin{array}{ll}
		 \mu^2(r-r_0)  - \displaystyle\frac{e^{-\mu r}}{r} & \mbox{if } r \leq r_0 \\
		 \\
		- \displaystyle\frac{e^{-\mu r}}{r} & \mbox{if } r > r_0
	\end{array}
    \right.~.
\end{equation}

From now on, the Nambu-type interaction potential, $V_{\rm N}(r)$, that shall be taken into consideration is the later given by (\ref{nambu-type-potential}) displayed in Fig.\ref{continuo}, and in the sequence we proceed to verify if it allows the existence of magnetic monopole-antimonopole bound states, the {\it monopolium}.

\section{Monopole-antimonopole bound states}
\begin{figure}[h]
\centering
\setlength{\unitlength}{1,0mm}
\includegraphics[width=6.05cm,height=5.5cm]{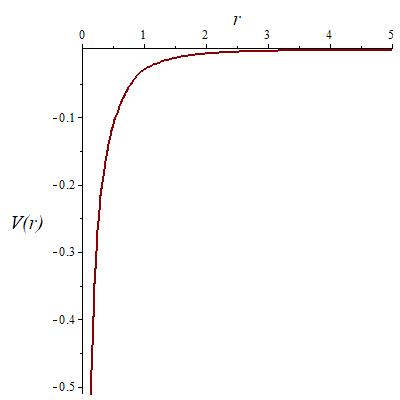}
\caption[]{The Nambu-type potential (\ref{nambu-type-potential}), after the {\it soldering} process at $r=r_0$, with $g=\mu=1$}
\label{continuo}
\end{figure}

In order to guarantee the existence of monopole-antimonopole bound states, the interaction potential must attend certain conditions. First of all the potential has to be attractive, nevertheless according to (\ref{nambu-type-potential}) the proposed Nambu-type potential is indeed attractive (see Fig.\ref{continuo}).

The attractiveness of the Nambu-type potential (\ref{nambu-type-potential}) is not a sufficient condition for the existence of monopole-antimonopole bound states. Furthermore, the interaction potential must fulfill the convergence condition \cite{convergence}:
\begin{equation}
\label{convergence}
\int\limits_0^\infty dr\,r \vert V_{\rm N}(r) \vert < \infty~,
\end{equation}
where, from the condition above (\ref{convergence}) together with (\ref{nambu-type-potential}), it stems that
\begin{eqnarray}
 \int\limits_0^\infty dr\,r \vert V_{\rm N}(r) \vert &\!\!=\!\!& \frac{g^2}{4\pi} \left[- \int\limits_0^{r_0} dr\,r \mu^2(r-r_0) + \int\limits_0^\infty dr\,r \frac{e^{-\mu r}}{r} \right] \nonumber\\
&\!\!=\!\!& \frac{g^2}{4\pi} \left(\frac{\mu^2 r_0^3}{6}+\frac{1}{\mu}\right) < \infty~.
\end{eqnarray}

Finally, the Nambu-type potential (\ref{nambu-type-potential}) must satisfy a last necessary bound state existing condition, the Set\^o-Bargmann bound \cite{convergence}, which provides the maximum number of bound states ($N_l$) associated to a particular angular momentum quantum number, $l$. Although the fulfillment of previous convergence condition (\ref{convergence}) allows the existence of bound states, it is the Set\^o-Bargmann bound \cite{convergence} that guarantees if there is at least one:
\begin{equation}
\label{seto-bargmann}
N_l \leq \left\lfloor{\frac{1}{2l+1} \int\limits_0^\infty dr\, r \vert V_{\rm N}(r) \vert}\right\rfloor~,
\end{equation}
with $\lfloor X \rfloor$ being the floor function of $X$ since the number of bound states ($N_l$) is a non-negative integer, which is given by
\begin{equation}
N_l \leq \left\lfloor{\frac{1}{2l+1}\frac{g^2}{4\pi} \left(\frac{\mu^2 r_0^3}{6}+\frac{1}{\mu}\right)}\right\rfloor~.
\end{equation}

It should be pointed out that the Set\^o-Bargmann condition (\ref{seto-bargmann}) also establishes a maximum value for the angular momentum quantum number ($l_{\rm m}$), inasmuch as $l=l_{\rm m}$ implies $N_{l_{\rm m}}=1$, so
\begin{equation}
l \leq l_{\rm m} = \left\lfloor \frac{1}{2} \left(\int\limits_0^\infty dr\, r \vert V_{\rm N}(r) \vert -1\right)\right\rfloor~.
\end{equation}
Hence, for the Nambu-type potential (\ref{nambu-type-potential}), the maximum angular momentum ($l_{\rm m}$) -- above which there is no bound states anymore -- reads
\begin{equation}
l \leq l_{\rm m} = \Biggl\lfloor \frac{1}{2} \left[\frac{g^2}{4\pi} \left(\frac{\mu^2 r_0^3}{6}+\frac{1}{\mu}\right) -1\right]\Biggr\rfloor~.
\end{equation}

In the sequel, owing to the fact that all the necessary conditions for monopole-antimonopole bound states have been satisfied by the proposed Nambu-type potential (\ref{nambu-type-potential}), we shall proceed the calculation of the {\it monopolium} ground state energy.

\section{The monopolium ground state energy}
In order to compute the {\it monopolium} ground state energy, associated to the central Nambu-type potential (\ref{nambu-type-potential}), we make use of the variational method, namely, we assume an isotropic wave function ansatz ($\psi(r)$) \cite{galindo-pascual} depending on a length-dimensional free parameter $(d)$ for posterior monopole-antimonopole hamiltonian vacuum expectation value ($\langle\hat{H}\rangle$) calculation followed by its minimization with respect to parameter $d$. It should be stressed that the ground state corresponds to the orbital $1s$ -- principal quantum number $n=1$, thereby orbital (angular momentum) quantum number $l=0$ -- implying unavoidably that the ground state wave function be spherically symmetric, thus $\psi(r)\equiv R_1(r)$. By taking all of this into consideration, the ansatz for the isotropic normalized wave function, is defined by:
\begin{equation}
\label{ansatz}
\psi(r)\equiv R_1(r)=\frac{1}{\sqrt{\pi d^3}}~e^{-\frac{r}{d}}~,
\end{equation}
where
\begin{equation}
\int d^3r\, |\psi(r)|^2 = 1~,
\end{equation}
with $d$ being the length-dimensional free parameter.

From the hamiltonian operator:
\begin{equation}
\label{hamiltonian}
\hat{H} = -\dfrac{1}{2m} \nabla^2 + V_{\rm N}(r)~,
\end{equation}
and the isotropic ground state wave function ansatz (\ref{ansatz}), we get
\begin{equation}
\hat{H} \psi(r) = -\frac{1}{2mr^2}\frac{\partial}{\partial r}\left(r^2 \frac{\partial}{\partial r}\right)\psi(r) + V_{\rm N}(r)\psi(r)~.
\end{equation}
In the following, let us calculate the ground state expectation value of the hamiltonian operator (\ref{hamiltonian}):
\begin{equation}
\label{mean-h}
\langle\hat{H}\rangle \equiv \langle\psi|\hat{H}|\psi\rangle = \int d^3r\, \psi^*(r)\hat{H}\psi(r)~,
\end{equation}
where it is convenient to remind that in (\ref{hamiltonian}) $m$ refers to the reduced mass of the monopole-antimonopole system. As a result, it stems from (\ref{mean-h}) that
\begin{eqnarray}
\langle\hat{H}\rangle &\!\!=\!\!& \frac{1}{2md} + \frac{ g^2}{\pi d^3}\left[\mu^2 e^{-\frac{2r_0}{d}}\left(-\frac{d^2r_0^2}{4}-\frac{d^3r_0}{2}-\frac{3d^4}{8}\right)\right]+ \nonumber\\
    &\!\!+\!\!&\frac{g^2}{\pi d^3}\left[\mu^2\left(\frac{3d^4}{8}-\frac{d^3r_0}{4}\right)-\frac{d^2}{(2+\mu d)^2}\right]~,
\end{eqnarray}
with $r_0$ being given by (\ref{r0}). Also, the mean radius of the {\it monopolium} bound state reads
\begin{equation}
\label{mean-r}
\langle r \rangle \equiv \langle\psi|r|\psi\rangle = \int d^3r\, \psi^*(r) r \psi(r)~,
\end{equation}
such that from (\ref{mean-r}):
\begin{equation}
    \langle r \rangle = \frac{3d}{2}~.
\end{equation}

Now, let us proceed the computation of the monopole-antimonopole ground state energy ($E_{\rm gs}$) and its corresponding characteristic radius ($\xi_{\rm gs}$), both depending on the {\it monopolium} reduced mass ($m$) and the interaction characteristic length ($\lambda=\frac{1}{\mu}$), namely the Compton wavelength associated to the quantum messenger particle with mass $\mu$. Accordingly:
\begin{equation}
E_{\rm gs} = \langle\hat{H}\rangle \big|_{d=d_{\rm m}} \aand
\xi_{\rm gs} = \langle r \rangle \big|_{d=d_{\rm m}}~,
\end{equation}
where $d_{\rm m}\equiv d_{\rm m}(m,\mu)$ stems from solving numerically\footnote{Use has been made of Maple software.}  the following transcendental equation:
\begin{equation}
\frac{\partial \langle\hat{H}\rangle}{\partial d} \Bigg|_{d=d_{\rm m}} = 0~,
\end{equation}
for a variety of {\it monopolium} reduced mass ($m$) and interaction characteristic length ($\lambda$). It is worthy to mention that we have used the Dirac charge quantization:
\begin{equation}
\label{dirac}
 eg = \frac{n}{2}~,~~ n \in \mathbb{Z}~,
\end{equation}
in order to establish the magnetic charge of the monopole (antimonopole), where $e$ stands for the electron charge. In addition, the lowest possible value for the monopole magnetic (positive) charge takes place when $n=1$, {\it e.g.} $g_{\rm D}\equiv\frac{1}{2e}$, so-called Dirac charge.

Finally, the numerical results for the ground state energy ($E_{\rm gs}$) and the characteristic radius ($\xi_{\rm gs}$) of the monopole-antimonopole bound state, calculated for a range of {\it monopolium} reduced mass ($m$) and mediating boson mass ($\mu$), are collected in Tab.\ref{tab:table1} and discussed throughout the next section.

\begin{table}[h!]
\begin{center}
\begin{tabular}{|c|c|c|c|c|}
\hline
$m$ (GeV) & $\mu$ (GeV) & $\lambda$ (fm) & $E_{\rm gs}$ (eV) & $\xi_{\rm gs}$ (fm) \\
\hline
\multirow{3}{*}{$10^{-1}$} & $4\cdot10^{-3}$ & $4,925\cdot10^1$  & $-6,483\cdot10^{0}$ & $6,126\cdot10^4$ \\
\cline{2-5}
& $2\cdot10^{-3}$ & $9,850\cdot10^1$ &  $-5,953\cdot10^{0}$ & $2,458\cdot10^5$ \\
\cline{2-5}
& $1\cdot10^{-3}$ & $1,970\cdot10^2$ &  $-1,569\cdot10^{0}$ & $9,849\cdot10^5$ \\
\hline
\multirow{3}{*}{$10^0$} & $4\cdot10^{-2}$ & $4,925\cdot10^0$  & $-6,483\cdot10^{1}$ & $6,126\cdot10^3$ \\
\cline{2-5}
& $2\cdot10^{-2}$ & $9,850\cdot10^0$ &  $-5,953\cdot10^{1}$ & $2,458\cdot10^4$ \\
\cline{2-5}
& $1\cdot10^{-2}$ & $1,970\cdot10^1$ & $-1,569\cdot10^{1}$ & $9,849\cdot10^4$ \\
\hline
\multirow{3}{*}{$10^1$} & $4\cdot10^{-1}$ & $4,925\cdot10^{-1}$  & $-6,483\cdot 10^{2}$ & $6,126 \cdot 10^2$ \\
\cline{2-5}
& $2\cdot10^{-1}$ & $9,850\cdot10^{-1}$  & $-5,953 \cdot 10^{2}$ & $2,458 \cdot 10^3$ \\
\cline{2-5}
& $1\cdot10^{-1}$ & $1,970\cdot 10^0$ & $-1,569 \cdot 10^{2}$ & $9,849 \cdot 10^{3}$ \\
\hline
\multirow{3}{*}{$10^2$} & $4\cdot10^{0}$ & $4,925\cdot10^{-2}$  & $-6,483\cdot 10^{3}$ & $6,126 \cdot 10^1$ \\
\cline{2-5}
& $2\cdot10^{0}$ & $9,850\cdot10^{-2}$ & $-5,953 \cdot 10^{3}$ & $2,458 \cdot 10^2$ \\
\cline{2-5}
& $1\cdot10^{0}$ & $1,970\cdot 10^{-1}$ & $-1,569 \cdot 10^{3}$ & $9,849 \cdot 10^{2}$ \\
\hline
\multirow{3}{*}{$10^3$} & $4\cdot10^{1}$ & $4,925\cdot10^{-3}$ &  $-6,483\cdot 10^{4}$ & $6,126 \cdot 10^0$ \\
\cline{2-5}
& $2\cdot10^{1}$ & $9,850\cdot10^{-3}$  & $-5,953 \cdot 10^{4}$ & $2,458 \cdot 10^1$ \\
\cline{2-5}
& $1\cdot10^{1}$ & $1,970\cdot 10^{-2}$ & $-1,569 \cdot 10^{4}$ & $9,849 \cdot 10^{1}$ \\
\hline
\multirow{3}{*}{$10^4$} & $4\cdot10^{2}$ & $4,925\cdot10^{-4}$ &  $-6,483\cdot 10^{5}$ & $6,126 \cdot 10^{-1}$ \\
\cline{2-5}
& $2\cdot10^{2}$ & $9,850\cdot10^{-4}$ & $-5,953 \cdot 10^{5}$ & $2,458 \cdot 10^0$ \\
\cline{2-5}
& $1\cdot10^{2}$ & $1,970\cdot 10^{-3}$ &  $-1,569 \cdot 10^{5}$ & $9,849 \cdot 10^{0}$ \\
\hline\multirow{3}{*}{$10^5$} & $4\cdot10^{3}$ & $4,925\cdot10^{-5}$ & $-6,483\cdot 10^{6}$ & $6,126 \cdot 10^{-2}$ \\
\cline{2-5}
& $2\cdot10^{3}$ & $9,850\cdot10^{-5}$ & $-5,953 \cdot 10^{6}$ & $2,458 \cdot 10^{-1}$ \\
\cline{2-5}
& $1\cdot10^{3}$ & $1,970\cdot 10^{-4}$ & $-1,569 \cdot 10^{6}$ & $9,849 \cdot 10^{-1}$ \\
\hline
\end{tabular}
\end{center}
\caption{\label{tab:table1} The ground state energy ($E_{\rm gs}$) and characteristic radius ($\xi_{\rm gs}$) for various values of {\it monopolium} reduced mass ($m$) and mediating boson mass ($\mu$), as well as the respective interaction characteristic length ($\lambda$).}
\end{table}

\section{Discussions}

Some interesting results can be observed from Tab.\ref{tab:table1}. We see that a wide range of masses are possible \cite{mass_range} for both the monopole-antimonopole reduced mass ($m$), from $10^{-1}$ to $10^5$GeV, and the mediating boson mass ($\mu$), from $10^{-3}$ to $10^3$GeV.



A closer look at Tab.\ref{tab:table1} leads to very interesting comparison. Remarkably, the ground state energy $E_{\rm gs}=-15,69$eV founded for the {\it monopolium} reduced mass, $m=10^0$GeV, and mediating boson mass $\mu=10^{-2}$GeV, are close to the value of the hydrogen atom ground state ($1s$ orbital), $E_{1s}=-13,61$eV. This allows the hypothesis of emissions from a Nambu-type {\it monopolium} atom close to the energy spectrum of visible light, giving hints about how to search for magnetic monopoles and, especially, for their bound states. It should be pointed out that, carrying out a fine tunning on the parameters $m$ and $\mu$ would lead to a {\it monopolium} ground state energy equal to the hydrogen atom, however this was not the main goal of this work.

Additionally, all those hydrogen-like ground state values of few eV (see Tab.\ref{tab:table1}) also justify the use of Schr\"odinger equation. Since the magnetic  monopoles interaction is orders in magnitude stronger than that between electric monopoles  carriers -- due to Dirac quantization condition, which gives a much higher value for the magnetic charge -- one would expect that the energy scale of this interaction should be in the relativistic regimen. Notwithstanding our expectancy, the use of Schr\"odinger equation instead of Dirac equation has been shown an appropriate choice, as verified for the cases where the ground state values are few order of magnitude in eV. However, for higher {\it monopolium} reduced mass and mediating boson mass the non relativistic approach might not be valid since the ground state energies enters the MeV scale. To reinforce the validity of the non-relativistic approach, we can estimate the root mean square of the monopole (antimonopole) velocity, $\sqrt{\langle v^2 \rangle}$, by invoking the Heisenberg uncertainty principle, ${\Delta r}{\Delta p}\geq \hbar/2$. Thanks to the isotropy of Nambu-type potential (\ref{nambu-type-potential}), it follows that:

\begin{equation}
\sqrt{\langle v^2 \rangle } \geq \frac{\hbar}{2m\xi_{\rm gs}}~,
\end{equation}
subsequently, from Tab.\ref{tab:table1}, by substituting for instance, $m=10^0$GeV and $\xi_{\rm gs}= 9,849 \cdot 10^4$fm -- which is the result close to the hydrogen atom ground state and mean radius -- we find $\sqrt{\langle v^2 \rangle} \approx 6\cdot 10^2$m/s. This assure us that our results are valid, leastwise for lower monopole-antimonopole reduced masses, nevertheless as it increases, the ground state energy ($E_{\rm gs}$) enters the MeV scale and the characteristic radius ($\xi_{\rm gs}$) reaches atomic nuclei scale, thence requiring an appropriate relativistic approach. As a next step it is worthwhile to consider the Darwin potential \cite{darwin} as a first order relativistic correction to the Nambu-type potential (\ref{nambu-type-potential}) and thereafter a full relativistic analysis via Dirac theory.


\section*{Acknowledgements}
The authors thank the anonymous referee for helpful comments. O.M.D.C. dedicates this work to his father (Oswaldo Del Cima, {\it in memoriam}), mother (Victoria M. Del Cima, {\it in memoriam}), daughter (Vittoria) and son (Enzo). Dedicated to the memory of Dr. C\'esar Augusto Parga Rodrigues (1939-2021). CAPES-Brazil is acknowledged for invaluable financial help.


\begin{references}

\bibitem{vaschy} A. Vaschy, ``Trait\'e d'\'electricit\'e et de magn\'etisme: Th\'eorie et applications, instruments et m\'ethodes de mesure \'electrique'', Baudry, Paris (1890).

\bibitem{curie} P. Curie, S\'eances de la Soci\'et\'e Fran\c aise de Physique (1894) 76.

\bibitem{dirac} P.A.M. Dirac, Proc. Roy. Soc. A133 (1931) 60.

\bibitem{anderson} C.D. Anderson, Science, 76 (1932) 238.

\bibitem{cf} N. Cabibbo and E. Ferrari, Nuovo Cim. 23 (1962) 1147.

\bibitem{thooft} G. 't Hooft, Nucl. Phys. B79 (1974) 276.

\bibitem{polyakov} A.M. Polyakov, JETP Lett. 20 (1974) 194.

\bibitem{polchinski} J. Polchinski,  Phys. Rev. Lett. 75 (1995) 4724.

\bibitem{schwinger} J. Schwinger, Science 165 (1969) 757.

\bibitem{nambu} Y. Nambu, Phys. Rev. D10 (1974) 12.

\bibitem{Volovik} G.E. Volovik, ``The Universe in a Helium Droplet'', Oxford University Press (2003).

\bibitem{Castelnovo} C. Castelnovo, R. Moessner and S.L. Sondhi, Nature 45 (2008) 42.

\bibitem{Mol2009} L.A. M\'{o}l, R.L. Silva, R.C. Silva, A.R. Pereira, W.A. Moura-Melo and B.V. Costa, J. Appl. Phys. 106 (2009) 063913.

\bibitem{Silva2013} R.C. Silva,  R.J. C Lopes, L.A.A. M\'{o}l, W.A. Moura-Melo, G.M. Wysin and A.R. Pereira, Phys. Rev. B87 (2013) 014414.

\bibitem{Morgan2011} J.P. Morgan, A. Stein, S. Langridge and C. Marrows, Nature Phys. 7 (2011) 75.

\bibitem{Marrows2019} S.A. Morley, J.M. Porro, A. Hrabec, M.C. Rosamond, D.A. Venero, E.H. Linfield, G. Burnell, Mi-Young Im, P. Fischer, S. Langridge, and C.H. Marrows, Scientific Reports 9 (2019) 15989.

\bibitem{Goryca} M. Goryca, X. Zhang, J. Li, A.L. Balk, J.D. Watts, C. Leighton, C. Nisoli, P. Schiffer and S.A. Crooker, Phys. Rev. X11 (2021) 011042.

\bibitem{macro} The MACRO Collaboration and M. Ambrosio et al., Eur. Phys. J. C25 (2002) 511.

\bibitem{balestra} S. Balestra, S. Cecchini, M. Cozzi et al., Eur. Phys. J. C55 (2008) 57.

\bibitem{epele2009} L.N. Epele, H. Fanchiotti, C.A. Garc\'ia Canal and V. Vento, Eur. Phys. J. C62 (2009) 587.

\bibitem{epele2012} L.N. Epele, H. Fanchiotti, C.A. Garc\'ia Canal, V.A. Mitsou and V. Vento, Eur. Phys. J. Plus 127 (2012) 60.

\bibitem{vento2021} V. Vento, Eur. Phys. J. C81 (2021) 229.
 
\bibitem{epele2008} L.N. Epele, H. Fanchiotti, C.A. Garc�a Canal and V. Vento,  Eur. Phys. J. C56 (2008) 87.

\bibitem{convergence}  K. Kodaira, Amer. Jour. Math. 71 (1949) 921; V. Bargmann, Proc. Nat. Acad. Sci. (USA) 38 (1952) 961; J. Schwinger, Proc. Nat. Acad. Sci. (USA) 47 (1960) 122; N. Set\^o, Publ. RIMS 9 (1974) 429.

\bibitem{galindo-pascual} A. Galindo and P. Pascual, ``Quantum Mechanics I'', Springer-Verlag (1990).

\bibitem{mass_range} L. Patrizii and M. Spurio, Ann. Rev. Nucl. Part. Sci. 65 (2015) 279.

\bibitem{darwin} D. Alba, H. Crater and L. Lusanna, Int. J. Mod. Phys. A16 (2001) 3365.


\end{references}
\end{document}